\documentclass[graybox]{editor/svmult}

\usepackage{type1cm}        % activate if the above 3 fonts are
\usepackage{makeidx}         % allows index generation
\usepackage{graphicx}        % standard LaTeX graphics tool
\usepackage{multicol}        % used for the two-column index
\usepackage[bottom]{footmisc}% places footnotes at page bottom

\usepackage{newtxtext}       % 
\usepackage[varvw]{newtxmath}       % selects Times Roman as basic font
\usepackage[numbers, round]{natbib}

\usepackage{siunitx}
\DeclareSIUnit{\permille}{\text{\textperthousand}}
\DeclareSIUnit{\molar}{M}
\DeclareSIUnit{\bar}{bar}

\makeindex             % used for the subject index
\begin{document}

\title*{Laser-based mass spectrometry for the detection of signatures of life within our Solar System }
\titlerunning{LIMS for Detecting Signatures of Life}
\author{Andreas Riedo \orcidID{0000-0001-9007-5791} and\\ 
Salome Gruchola \orcidID{0000-0002-9757-1402} and\\
Nikita J. Boeren \orcidID{0000-0001-6162-6953} and\\ 
Peter Keresztes Schmidt \orcidID{0000-0002-4519-8861} and\\
Luca N. Knecht \orcidID{0009-0007-2548-6501} and\\
Youcef Sellam \orcidID{0009-0002-7690-5386} and\\
Marek Tulej \orcidID{0000-0001-9823-6510} and\\
Peter Wurz \orcidID{0000-0001-9823-6510}
}
\authorrunning{LIMS for Detecting Signatures of Life}

\institute{Andreas Riedo, andreas.riedo@unibe.ch.\\
Space Research and Planetary Sciences, Physics Institute, University of Bern, Bern, Switzerland. \\
NCCR PlanetS, University of Bern, Bern, Switzerland }
\maketitle

%\abstract*{Each chapter should be preceded by an abstract (no more than 200 words) that summarizes the content. The abstract will appear \textit{online} at \url{www.SpringerLink.com} and be available with unrestricted access. This allows unregistered users to read the abstract as a teaser for the complete chapter.
%Please use the 'starred' version of the \texttt{abstract} command for typesetting the text of the online abstracts (cf. source file of this chapter template \texttt{abstract}) and include them with the source files of your manuscript. Use the plain \texttt{abstract} command if the abstract is also to appear in the printed version of the book.}

\abstract{The search for signatures of life beyond Earth has been a major goal of space research and astrobiology for decades. The combination of expanded knowledge on Solar System bodies from past missions and advancements in in-situ detection technologies may place humanity on the verge of discovering extraterrestrial life. Here, we highlight the current measurement capabilities of Laser Ionisation Mass Spectrometry for the detection of several classes of signatures of life of high relevance to current astrobiology-focused missions. This includes the detection of microstructures within complex geological hosts by chemical depth profiling, sulphur isotope fractionation signatures, and the detection of various classes of organic molecules. The recorded mass spectrometric data can be fed into network and machine learning analysis routines, which are powerful tools for the unbiased detection of signatures of life, including agnostic detection of biosignatures. We demonstrate that Laser Ionisation Mass Spectrometry is a novel and promising technology for future application on space exploration missions devoted to life detection.}

\section{Introduction}
\label{sec:intro}

Mass spectrometry is one of the most sensitive and analytically most specific technique for the chemical composition analysis of any solid. As a result, mass spectrometers are typically part of the instrument payload on space exploration missions, either for in-situ chemical analyses of a planetary atmosphere or on the planetary surface to search for signatures of life. The latter dates back to the 1970s, when mankind sent the two Viking landers to the surface of Mars \citep{Soffen1976} to detect signatures of life using Gas Chromatography-Mass Spectrometry (GC-MS) \citep{Oyama1972}, unfortunately without providing conclusive evidence for the presence of extinct or extant life. GC-MS instrumentation for space exploration has been continuously further developed in terms of measurement capabilities, and one of the most sophisticated GC-MS systems is the Sample Analysis at Mars (SAM) instrument \citep{Mahaffy2012}, which is part of NASA’s Curiosity rover on Mars \citep{Grotzinger2012}. Although GC-MS instruments belong to the state-of-the-art equipment in laboratory research, they experienced limitations when applied on planetary surfaces, e.g., in the presence of chlorinated species \citep{Royle2018}. 

In this chapter, we review the current measurement capabilities of two Laser Ionisation Mass Spectrometry (LIMS) systems developed in-house and applied to the detection and identification of signatures of life. Previously we have shown that LIMS has the potential to identify micrometre-scale features within a complex geological host \citep{Tulej2015, Sellam2025}, to provide isotope ratio measurements with per mille accuracy in delta notation to identify possible fractionation processes \citep{Riedo2021}, and to detect organic molecules important for life as we know it, including amino acids \citep{Ligterink2020}, lipids \citep{Boeren2022}, and nucleobases \citep{Boeren2025}, at surface concentrations down to \unit{\femto\mol\per\square\milli\metre}. The latter is of great interest for current space science, as the mass spectrometric system already fulfils many measurement requirements, e.g., for a landed mission on Enceladus \citep{MacKenzie2021} or Europa \citep{Hand2017}.

The NCCR PlanetS allowed to extend and better understand the measurement capabilities of one of the LIMS systems, which is specifically designed and used for the detection and identification of organic molecules. Initially, studies were conducted solely on amino acids. Thanks to NCCR PlanetS, in-depth studies of lipids and nucleobases were made possible, opening new doors in this competitive field of space science.

\section{Laser Ionisation Mass Spectrometry}
\label{sec:LIMS}

\begin{figure}[htbp]
\sidecaption[t]
\includegraphics[width = 6 cm]{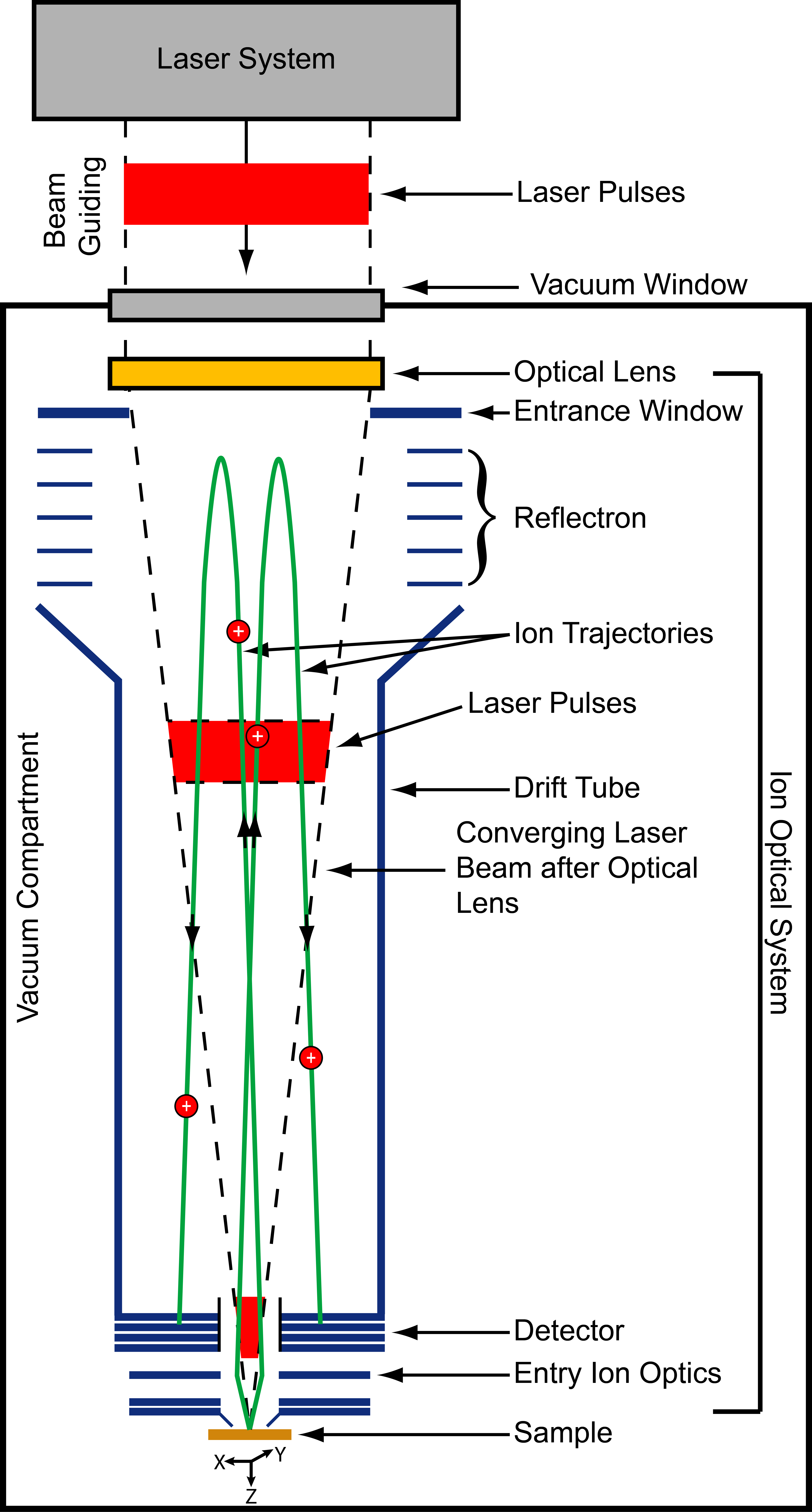}
\caption{Schematics and operating principles of our miniature LIMS setups.}
\label{fig:1}     
\end{figure}

For more than two decades, the laser mass spectrometry group at the University of Bern has been developing and further optimising LIMS systems \citep{Rohner2003, Riedo2013a, Wiesendanger2019} and new measurement methodologies for the chemical composition analysis of various samples of interest \citep{Boeren2025, Neubeck2015, Wiesendanger2018, Grimaudo2019, Grimaudo2020}. In the following, we briefly describe the schematics (Fig. \ref{fig:1}) and operating principles of two of our miniature LIMS space prototype systems. We specifically apply these instruments to develop new measurement protocols for the detection of life signatures presented in this contribution \citep{Riedo2021, Gruchola2024, Riedo2025}.

The two LIMS systems both consist of the same core component, a reflectron-type time-of-flight (RTOF) mass analyser (\qty{160}{\milli\metre} x Ø \qty{60}{\milli\metre}), which can be coupled to various laser ablation or laser desorption ion sources (from femtosecond pulsed laser systems operating at \qty{775}{\nano\metre} down to \qty{258}{\nano\metre}, to nanosecond pulsed laser systems typically operating at \qty{266}{\nano\metre}). The main difference between these two modes of operation is the laser irradiance applied onto the sample surface. For element and isotope composition analyses, laser ablation conditions are required with laser irradiances in the \qty{e9}{\watt\per\square\centi\metre} to \qty{e12}{\watt\per\square\centi\metre} range (ns to fs laser pulses), resulting in efficient removal and atomisation of material from a solid with each laser pulse fired. For the detection of organic molecules, laser irradiances at the level of \qty{e6}{\watt\per\square\centi\metre} have to be applied, which allow for the gentle removal and ionisation of molecules from the surface without severe fragmentation, if any. We typically have one LIMS system operating under ablation conditions (referred to as the Laser Mass Spectrometer (LMS) \citep{Sellam2025, Riedo2013b}) and one system operating under desorption conditions (referred to as ORganics Information Gathering INstrument (ORIGIN) \citep{Boeren2025}). In both LIMS setups, the laser system and beam delivery optics are located outside a vacuum chamber for simplicity, while the mass analyser is operated inside a vacuum chamber. A lens system focusses the laser beam through the mass analyser’s central axis onto the sample surface to spot sizes ranging from about \qtyrange{10}{30}{\micro\metre} in diameter. A larger spot size is preferred for laser desorption studies, as it reduces the laser irradiances, while for laser ablation studies a smaller spot size allows for higher spatial resolution. The sample of interest is placed close to the entrance of the ion optical system of the mass analyser, typically within mm-range. For each laser pulse-material interaction, a distinct layer of sample material is removed and partially ionised. The positively charged ions can enter the ion optical system of the mass analyser, where they are confined and accelerated towards the field-free drift tube. The species are then redirected by the reflectron towards the detector system \citep{Riedo2017} by passing through the drift tube a second time. The ions arrive at the detector system sequentially according to their mass-to-charge ratio \citep{Becker2007}. Read-out electronics are used to record the electrical signal generated by the detector system. We typically record the detector signal up to \qty{20}{\micro\second} after the laser pulse hits the surface, resulting in the TOF spectrum covering up to about m/z 700.

\subsection{Measurement Procedures}
\label{sec:MeasProc}

The measurement procedure depends on the scientific question and the sample under investigation. Therefore, there is no typical LIMS measurement routine that we follow one to one. Although the measurement protocols for both laser ablation and desorption studies are very similar, there are important differences between them, which are discussed below.

Laser ablation allows for the removal of a distinct layer of material from the sample under investigation with a lateral resolution of about \qtyrange{10}{15}{\micro\metre} (limited by the integrated optics) and vertical resolution of about nm (depending on the applied laser irradiance) \citep{Grimaudo2020, Grimaudo2015}. Therefore, by aiming the laser beam at a single surface position, a one-dimensional chemical depth profile can be obtained from successive laser pulses. By extending this procedure to multiple surface positions, 3D element maps of the sample can be derived, see e.g., \citep{Grimaudo2017}. The number of laser shots applied to a single surface position ranges from a few tens to a few thousand shots, depending on the scientific question and the sample. Since a TOF spectrum can be recorded for every single laser shot, the total number of recorded spectra can easily exceed one hundred thousand. If such high-resolution chemical information is not required, the acquisition system allows for on-card histogramming of spectra in such a way that one TOF spectrum stored on the host computer corresponds to histogrammed data obtained from, for example, 100 individual laser shots.

The first laser desorption study we conducted dates back to 2010 and was performed on a carbonaceous film using LMS coupled to a nanosecond pulsed laser system (laser wavelength $\lambda = \qty{266}{\nano\metre}$) \citep{Riedo2010}. This initial study was followed by studies on e.g., glycine and toluene, again using LMS, but this time coupled to a femtosecond laser system \citep{Tulej2014, Moreno2016}. Currently, we employ the ORIGIN setup with a nanosecond laser for desorption studies. Typically, we drop cast \qty{1}{\micro\litre} of sample solution into a sample cavity (\qty{0.2}{\milli\metre} x Ø \qty{3}{\milli\metre}) of a sample holder (e.g., a stainless-steel substrate) and let the solvent evaporate inside a clean bench, leaving a residual organic layer. The sample holder is precleaned through sonication in solvents, followed by argon ion sputtering; see Boeren et al., 2025 \citep{Boeren2025} for more details.

Gentle removal of organic molecules with little fragmentation (if present) is of primary interest in laser desorption studies. Pulse energies (laser wavelength $\lambda = \qty{266}{\nano\metre}$) at the level of few \unit{\micro\joule} only (measured at the sample surface) have been applied for such desorption studies, and generally 100 laser shots have been fired at a single surface position. Typically, we raster spot-wise over 40 single surface positions to compensate for the (potentially) inhomogeneous sample distribution within the sample cavity (see e.g., Fig. 5 in \citep{Boeren2025}). The number of surface positions or the number of laser shots per position can be easily varied \citep{Boeren2025}. In laser desorption studies, a TOF spectrum is recorded for each laser shot fired at the sample. Each TOF spectrum is screened for the presence of mass spectrometric peaks, and if present histogrammed into a single TOF spectrum.

\section{Promising Signatures of Life}
The Mars Science Definition team highlighted six groups of promising signatures of life that are highly relevant to current space exploration and astrobiology. These include organic molecules, isotope fractionation, micro and macro structures, fine-scale chemistry, and mineralogy \citep{Hays2017}. The following sections aim to highlight the versatility of LIMS in detecting biosignatures in different materials; of these six promising groups, LIMS has shown its potential in four, including organics, isotope fractionation, microstructures, and mineralogy. 

\begin{figure}[b]
\sidecaption
\includegraphics[width = 7 cm]{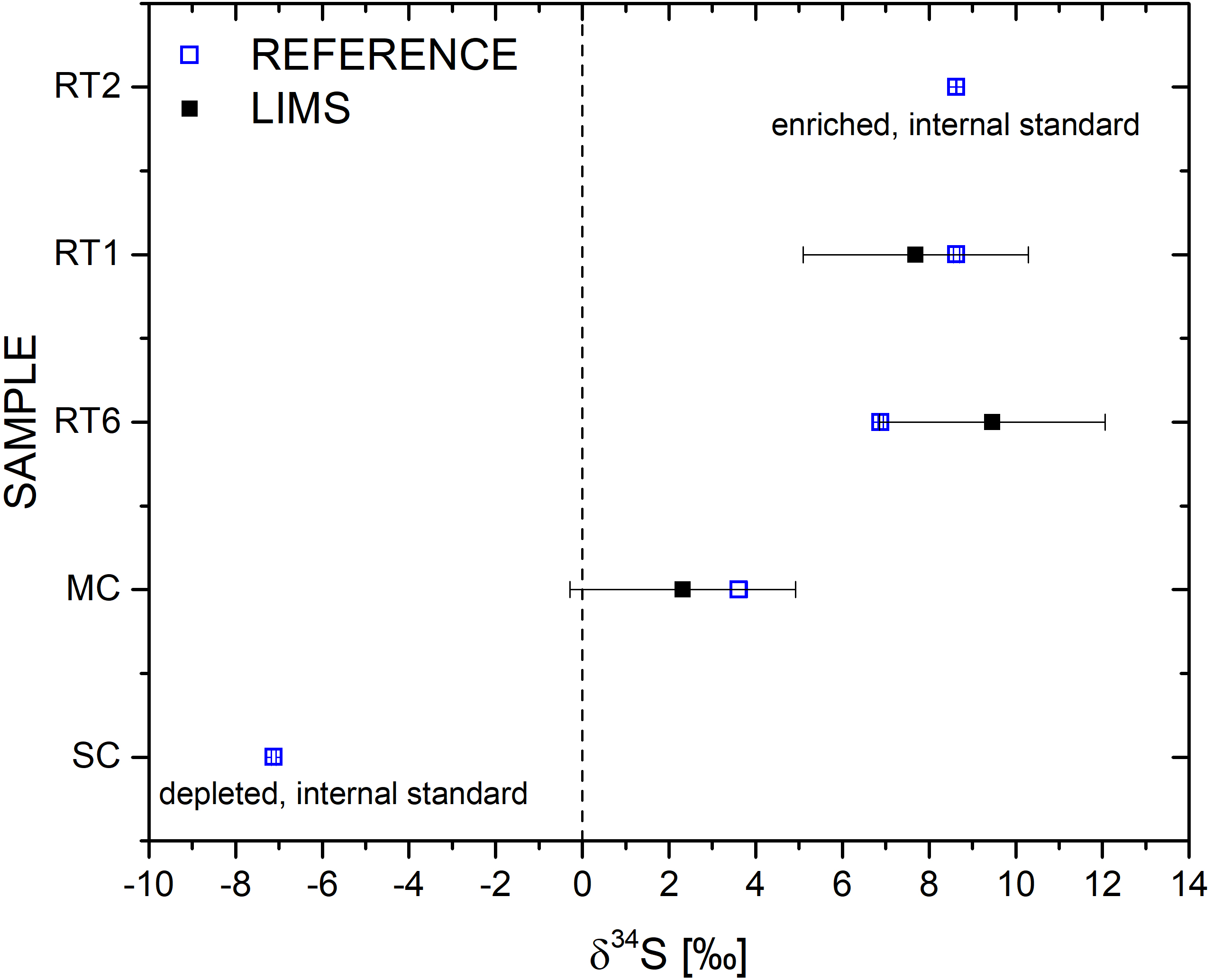}
\caption{Sulphur isotope measurements of five sulphur containing samples using LMS. The most enriched and depleted samples were used as internal calibration standards, allowing to quantify the remaining three samples within accuracies of $\delta ^{34}\text{S} = \qtyrange[range-phrase = -]{2}{3}{\permille}$. Figure taken from Riedo et al., 2021 \citep{Riedo2021}, licensed under CC-BY 4.0.}
\label{fig:2}     
\end{figure}

\subsection{Sulphur Isotope Fractionation}

Isotope fractionation is a very robust biosignature as it is less sensitive to environmental parameters, such as harsh temperature or ionisation conditions typically prevailing on Solar System bodies. Sulphur isotopes are of particular interest to various scientific fields. For example, the chemical analysis of sulphur-containing solids has contributed to a better understanding of geochemical processes that shaped our Earth’s atmosphere and crust \citep {Farquhar2002, Farquhar2003}, and has improved our understanding of Mars’s climate history and past and current habitability \citep {King2010, Ding2015}. For life detection on planetary objects, sulphur isotope fractionation is highly relevant as well \citep{Chela2018}. On Earth, organisms, such as sulphur reducing bacteria, exist that can fractionate sulphur isotopes up to values of $\delta ^{34}\text{S} = \qty{-70}{\permille}$ \citep{Wortmann2001}, which is significantly above fractionation achievable through physical or chemical processes, up to about $\delta ^{34}\text{S} = \qty{-20}{\permille}$ \citep{Chela2006}. Therefore, detecting isotope fractionation significantly above $\delta ^{34}\text{S} = \qty{-20}{\permille}$ is indicative of the existence of past or present life. However, no space instrument has demonstrated achieving such detection sensitivities and measurement capabilities so far.

In Riedo et al., 2021 we presented a novel measurement protocol that allows for sulphur isotope fractionation measurements using LMS with an estimated accuracy of about $\delta ^{34}\text{S} = \qtyrange[range-phrase = -]{2}{3}{\permille}$. A peak-like trend between applied pulse energy and recorded $^{34}$S/$^{32}$S signal was observed for the sulphur-containing samples (see Fig. 6 in \citep{Riedo2021}). The most enriched and depleted samples in $\delta ^{34}\text{S}$ were used as internal standards and allowed to quantify the fractionation values of the remaining three samples with accuracies of $\delta ^{34}\text{S} = \qtyrange[range-phrase = -]{2}{3}{\permille}$ (Fig. \ref{fig:2}). This is sufficiently accurate to distinguish between e.g., the $\delta ^{34}\text{S} = \qty{-20}{\permille}$ and $\delta ^{34}\text{S} = \qty{-70}{\permille}$ case. In said study, five different sulphur-containing samples, collected from three different field sites mimicking some parameters of past and current Mars, have been investigated. The samples differ in sulphur abundances (from \qtyrange{5.7}{96.5}{wt\percent}) and fractionation level (from $\delta ^{34}\text{S} = \qtyrange[retain-explicit-plus]{-7.12}{+8.63}{\permille}$), both measured with state-of-the-art laboratory techniques, as well as chemical composition. This measurement capability offers new perspectives for e.g., in-situ application on Mars, where sulphur deposits have been identified with sulphur abundances of up to about \qty{37}{wt\percent} (about \qty{6}{wt\percent} on average) \citep{King2010, Ding2015}.

\subsection{Fossil Structure Detection}

\begin{figure}[t]
\centering
\includegraphics[width = 11.7 cm]{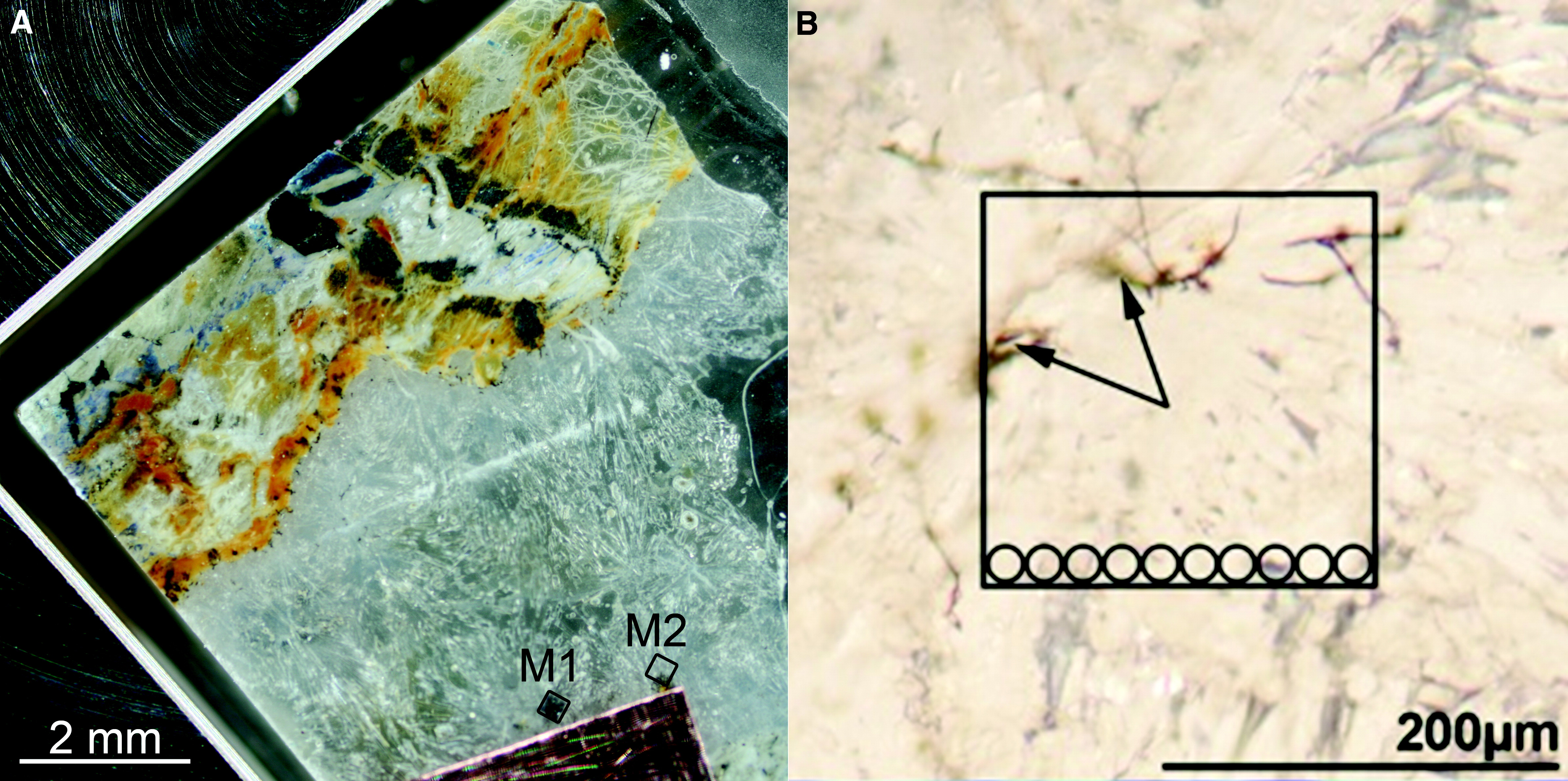}
\caption{A: Microscope images of the aragonite host showing two survey areas (M1 and M2). B: Area M1 showing micrometer-sized fossil structures in more detail. Figure adapted from Tulej et al., 2015.}
\label{fig:3}     
\end{figure}

The application of laser ablation conditions using LMS allows a spatially resolved visualisation of the chemical composition distribution within a solid. This measurement protocol was applied to an aragonite host (collected from the slow-spreading mid-Atlantic ridge \citep{Bach2011}) for the localisation and chemical composition analysis of micrometre sized fossil structures \citep{Tulej2015}. Fig. \ref{fig:3} shows the thin section of the aragonite host fixed onto a sample holder (left), containing micrometre-sized fossil structures (diameter of about \qtyrange{2}{5}{\micro\metre} and a length from about \qty{20}{\micro\metre} up to more than \qty{100}{\micro\metre}, right). Two areas on this aragonite host (M1 and M2), each \qtyproduct{200 x 200}{\micro\metre}, were chemically analysed on a raster of \numproduct{10 x 10} individual locations.

\begin{figure}[b]
\centering
\includegraphics[width = 11.7 cm]{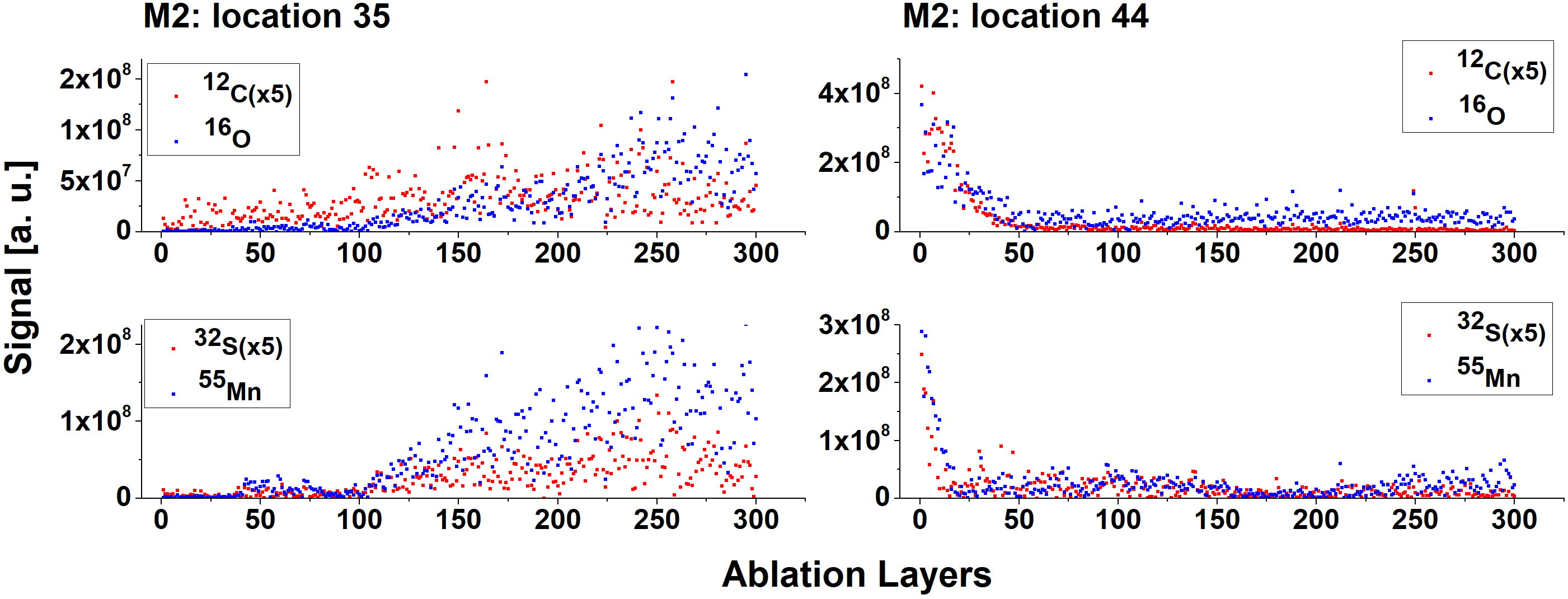}
\caption{Chemical depth profiles of locations \texttt{\#35} and \texttt{\#44} from measurement area M2 (see Fig. \ref{fig:3}). Through element profiling of biologically relevant elements the location of the fossil structures can be identified. Figure adapted from Tulej et al., 2015.}
\label{fig:4}     
\end{figure}

The chemical depth profiling procedure allows the localisation of fossil structures within the host by monitoring biologically relevant elements with increasing depth. Fig. \ref{fig:4} shows such depth profiles of two surface positions in M2 for the elements C, O, S, and Mn. The increased element intensities observed at the beginning of the depth profile at surface location \texttt{\#44} indicate a fossil structure at the surface, in contrast to the structure observed at location \texttt{\#35}. At a depth of the ablation layer around 100, a similar element abundance increase is observed, indicating a structure embedded within the aragonite host. As all mass spectrometric information of these ablation layers is stored, the specific mass spectra can be checked for further chemical analysis (see e.g., Fig. 6 or Table 2 in Tulej et al., 2015).

Another sample containing fossil structures analysed with LMS is a Messinian gypsum from Algeria \citep{Sellam2025}. The sample was retrieved from the Sidi Boutbal quarry (SB) in the Oran district, where extensive gypsum formations can be found. These deposits formed 5.97–5.33 Ma ago during the Messinian Salinity Crisis (MSC), during which the Mediterranean Sea was partially to completely isolated from the Atlantic Ocean, resulting in a substantial evaporation of the volume of the Mediterranean Sea \citep{Krijgsman1999, Hsue1973, Barker2018, Schopf2012, Roveri2014}. This increased the salinity of the seawater above contemporary levels, transforming it into a large hypersaline depositional environment, leading to the extermination of most eukaryotic life thriving in the Mediterranean Sea \citep{Bellanca2001}. The fossils preserved in Messinian gypsum therefore belong to a limited number of extremophile prokaryotes \citep{Allwood2013}. 

\begin{figure}[t]
\centering
\includegraphics[width = 11.7 cm]{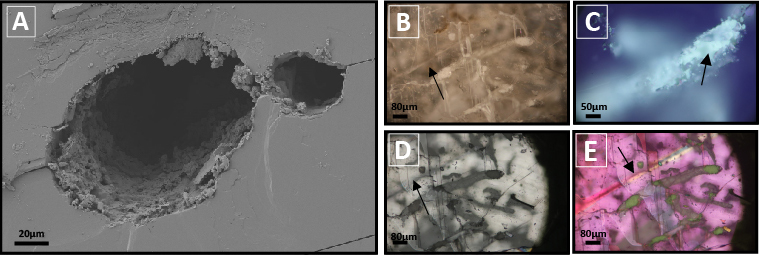}
\caption{Optical microscopy and SEM images of the Messinian gypsum sample studied with LMS. For full figure including caption see \citep{Sellam2025}. Figure adapted from \citep{Sellam2025}, licensed under CC-BY 4.0.}
\label{fig:5}     
\end{figure}

A fossil filament preserved within primary bottom-grown gypsum selenite formed during the MSC in the Chelif marginal basin, Algeria, was investigated using various measurement techniques, including LIMS \citep{Sellam2025}. Optical and SEM images of the studied sample are summarised in Fig. \ref{fig:5}. Through the investigation of the sample’s morphology, element composition, and mineralogy, the biogenicity and syngeneity of the fossil filament could be assessed.

\begin{figure}[b]
\centering
\includegraphics[width = 11.7 cm]{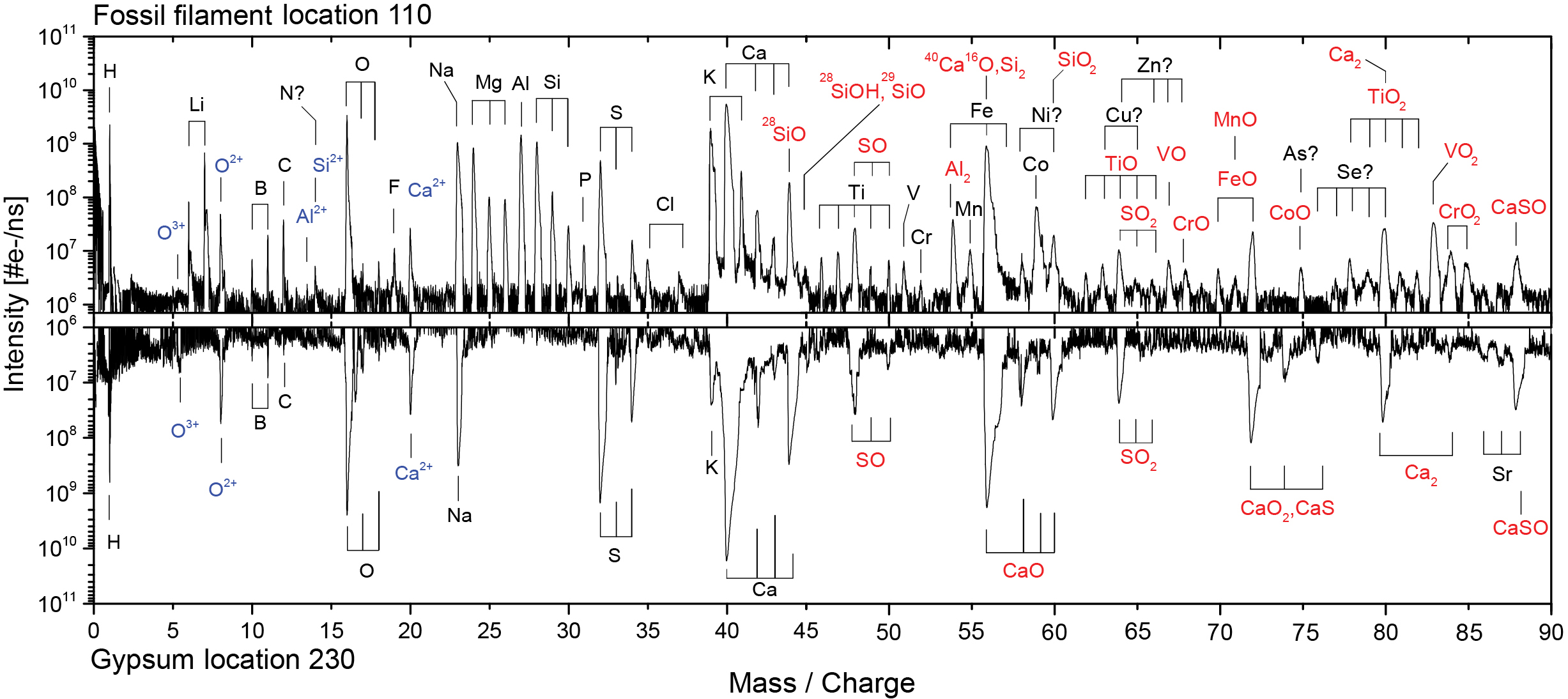}
\caption{Comparison of mass spectra recorded from the fossil filament (location 110, upper trace) and the gypsum host (location 230, lower trace, inverted). Figure adapted from \citep{Sellam2025}, licensed under CC-BY 4.0.}
\label{fig:6}     
\end{figure}

The area shown in Fig. \ref{fig:5}A, was subjected to a LMS raster analysis, acquiring depth profiles from 425 locations, which allowed to understand the spatial heterogeneity of the sample at the \unit{\micro\metre} spatial scale. A mass spectrum acquired from the gypsum (location 230) is compared to a mass spectrum of the fossil filament (location 110) in Fig. \ref{fig:6}. The mass spectrum of the filament reveals the presence of numerous metallic and non-metallic minor elements, including Li, B, C, F, Al, P, S, Cl, K, Ti, V, Cr, Mn, Fe, Co, and Sr, in addition to the major elements O, Mg, Si, and Ca. This contrasts the elements found in the mass spectrum of the gypsum, majorly H, O, S, Ca, and Na, with minor B, C, K, and Sr.

Messinian gypsum is a Mars analogue, as hydrated sulphate deposits have been detected on Mars. Therefore, this study has high relevance to the search for past life in Martian hydrated gypsum and sulphate deposits. Hydrated sulphates serve as archives of biological history on Earth and could potentially do the same on Mars. Future astrobiological investigations on Mars should therefore consider hydrated sulphate deposits as targets to search for signatures of past life.

\subsection{Single Cell Microbe Detection}

Chemical depth profiling can identify fossil structures within the sample of interest, and it also allows for the localisation and identification of single cells within a host material too. Riedo et al., 2020 \citep{Riedo2020} showed that LIMS can detect single cell microbes in a Mars analogue sample. In this study, 14 Martian mudstone analogues were examined, 7 samples were inoculated with microbes (Bacillus subtilis) at a cell density of about $10^6$  cm\textsuperscript{-3}; the other 7 samples were used as negative controls. Through chemical depth profiling carbon-rich layers have been identified within biotic and abiotic samples (see Fig. 4 in \citep{Riedo2020}). With all the mass spectrometry data available, these particular mass spectra can be investigated for further chemical information.

\begin{figure}[htbp]
\centering
\includegraphics[width = 11.7 cm]{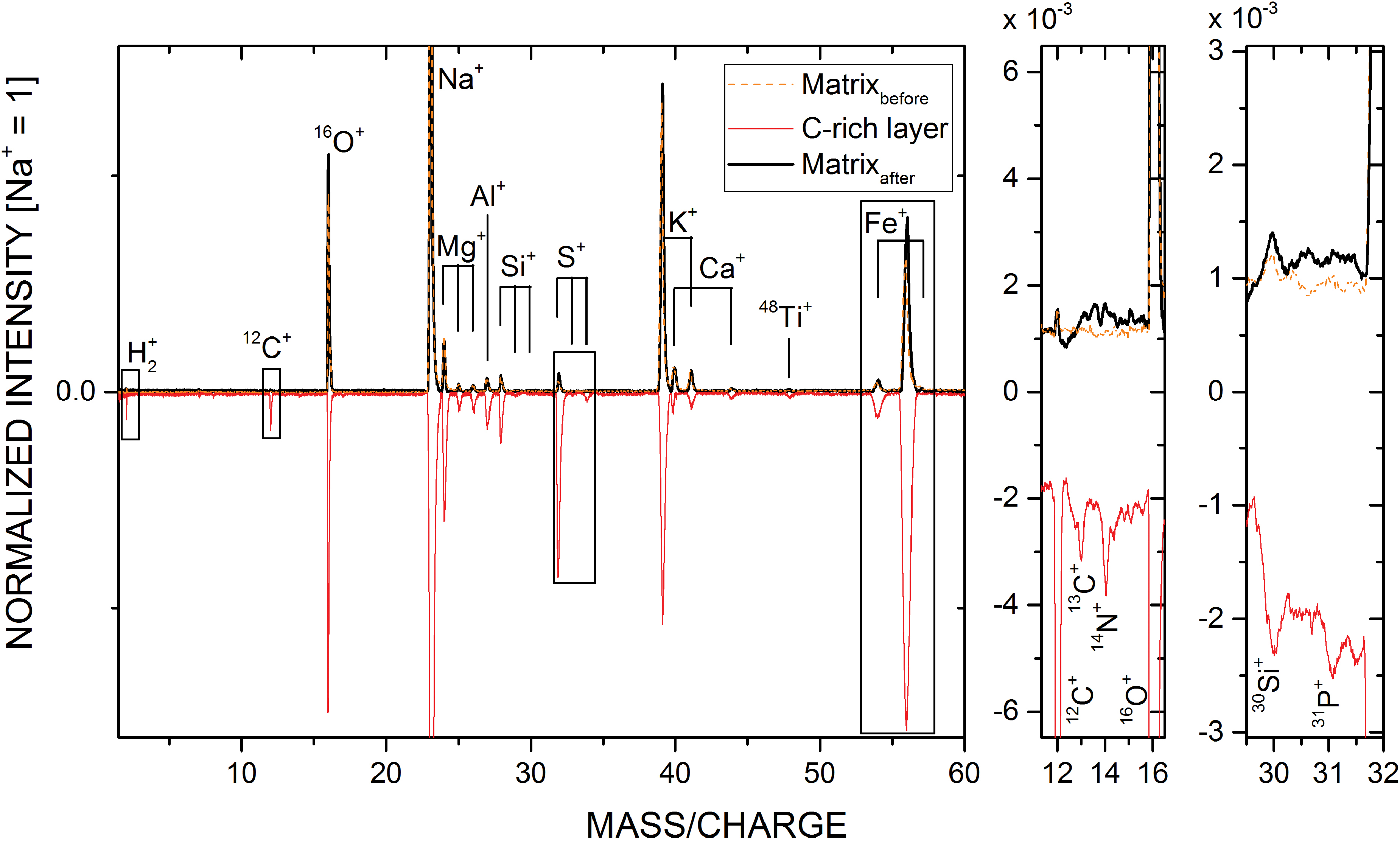}
\caption{LMS measurement of a Martian mudstone analogue sample artificially inoculated with cells. Compared to the matrix surrounding the carbon layer, a signal increase of the biologically relevant elements can be observed. Figure taken from \citep{Riedo2020}.}
\label{fig:7}     
\end{figure}

In Fig. \ref{fig:7} LMS measurements of a biotic sample material are shown. Three mass spectra are displayed; two pointing upwards, representing the chemical composition of the matrix measured before and after the carbon-rich layer identified by chemical depth profiling, and one pointing downwards, representing the chemistry of the carbon layer. As can be seen, the carbon signal is significantly enhanced compared to the surrounding matrix. In addition, an increased signal of the major and minor biologically relevant elements H, N, P, S, and Fe can be identified which is absent for the abiotic (no cell inoculation) cases (see Fig. 5 in \citep{Riedo2020}). Five biotic cases were identified within the study, which is at the same theoretical expectation level of two detections (amount of material removed during laser ablation). Therefore, this LMS protocol would not only allow for the identification of cells, but also for the differentiation between biotic and abiotic cases, which is of great interest for current astrobiology.

\subsection{Organic Molecule Detection}

Organic compounds are a fundamental class of biosignatures and targets for proposed space missions to Mars, Europa, Enceladus, and others \citep{Hand2017, MacKenzie2020, Goesmann2017, NASA2023}. In particular, molecular building blocks of life as we know it, such as amino acids, lipids, and nucleobases, are often highlighted, but the presence of a single compound is unlikely to be sufficient to confirm past or present life, as some have also been found in meteorites \citep{Sephton2005, Hand2017}. It is therefore crucial for instrumentation to be capable of identifying a broader range of molecular biosignatures in specific ratios and concentrations.

ORIGIN has been tested for its ability to detect various organic compounds without the need for pretreatment steps such as derivatisation. Studies have successfully identified amino acids \citep{Ligterink2020}, nucleobases \citep{Boeren2025}, lipids \citep{Boeren2022}, and polycyclic aromatic hydrocarbons (PAHs) \citep{Kipfer2022}. Fig. 8 presents selected mass spectra from these compound classes, showcasing distinct fragmentation patterns that facilitate the development of a spectral library for compound identification in natural samples. Molecular ions are marked with an asterisk next to their specified mass. Notably, this wide range of compounds was identified using a single instrument and measurement protocol.

\begin{figure}[t]
\centering
\includegraphics[width = 11.7 cm]{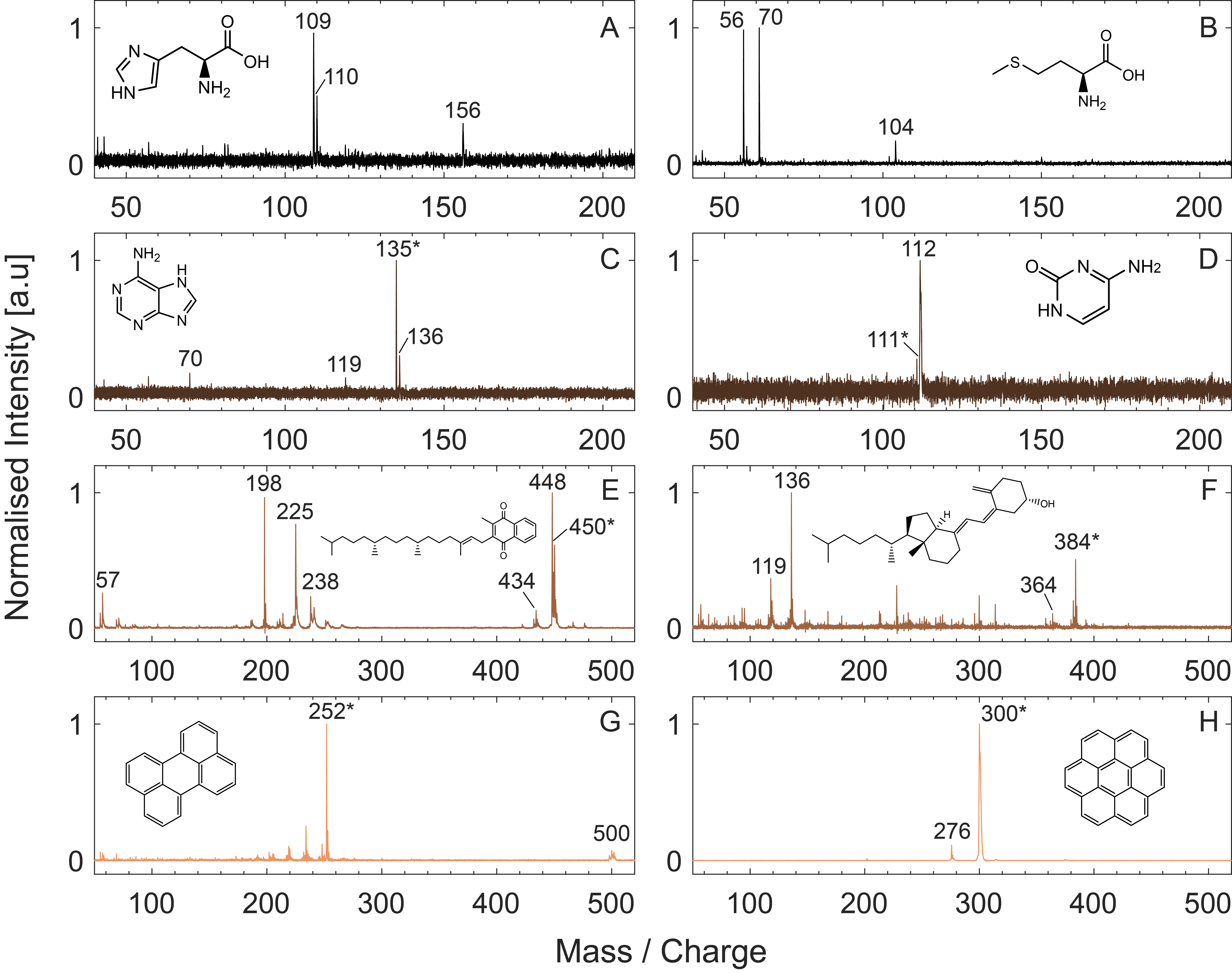}
\caption{Mass spectra recorded with ORIGIN of amino acids L-histidine (A) and L-methionine (B), nucleobases adenine (C) and cytosine (D), lipids phylloquinone (E) and cholecalciferol (F), and PAHs perylene (G) and coronene (H).}
\label{fig:8}     
\end{figure}

Twenty biotic and abiotic amino acids have been investigated, with the top spectra in Fig. \ref{fig:8} illustrating the mass spectra of histidine and methionine, displaying characteristic mass peaks that facilitate confident identification \citep{Ligterink2020}. Furthermore, amino acid mixtures were analysed in the presence of NaCl to assess potential interference from salts. The results confirmed that ORIGIN is unaffected by the presence of NaCl, a salt expected at percent-level concentrations on Ocean Worlds.

Boeren et al. 2025 examined six nucleobases, which are the core components of nucleic acids. All six, adenine, cytosine, 5-methylcytosine, guanine, thymine, and uracil, were readily identified, with mass spectra exhibiting parent ions and minimal fragmentation. Fig. \ref{fig:8}C and D display mass spectra of adenine and cytosine, in which the limited fragmentation is clearly observed due to their stable aromatic ring structures. 

Lipids are a diverse class of biomolecules and six representative prenol and sterol lipids were investigated, including cholecalciferol, $\alpha$-tocopherol, phylloquinone, menadione, 17$\alpha$-ethynylestradiol, and retinol \citep{Boeren2022}. Fig. \ref{fig:8}E and F display the mass spectra of phylloquinone and cholecalciferol, demonstrating their unique fragmentation patterns characterised by distinct parent and minor fragment peaks. Furthermore, the reliable detection and identification of mixtures containing amino acids, PAHs, and lipids was demonstrated, representing a milestone in the application of this detection technology.

Four PAHs (pyrene, perylene, anthracene, and coronene) were analysed following an identical measurement protocol \citep{Kipfer2022}. Fig. \ref{fig:8}G and H present the mass spectra of perylene and coronene, clearly showing intense molecular ion signals with minimal fragmentation of their aromatic ring structures.

Both qualitative and quantitative analyses of biosignatures are essential, as molecular abundance patterns can differentiate abiotic from biotic sources (e.g., lipid distributions as a function of carbon chain length) \citep{Georgiou2014}. ORIGIN enables direct quantification by correlating ion detection with peak area. As these measurements are based on residue left after drying a liquid sample, average surface concentrations are typically used for quantification. Linear correlation between the compound peaks and their surface concentration have been investigated and confirmed for amino acids, lipids, and PAHs \citep{Ligterink2020, Boeren2022, Kipfer2022}. Investigated concentration ranges spanned from a few \unit{\femto\mol\per\square\milli\metre} to \qty{100}{\pico\mol\per\square\milli\metre}, demonstrating ORIGIN’s high dynamic range and suitability for quantitative analyses. The results showed that order of magnitude quantitative information can be derived and over four orders of magnitude were covered with a single measurement procedure, setup, and settings, without requiring any detector adjustments. 

From the average surface concentration and the signal-to-noise ratio (SNR, with SNR $\geq$ 3$\sigma$), we calculate the limits of detection (LOD). For ORIGIN we determined a $\text{LOD}_{3\sigma}$ of \qty{52}{\femto\mol\per\square\milli\metre} for the nucleobase adenine (\qty{\sim 700}{\femto\mol\per\square\milli\metre} deposited from a \qty{1}{\micro\litre} aliquot of a \qty{5}{\micro\molar} solution). For ExoMars, the MOMA Laser Desorption-Mass Spectrometry (LD-MS) instrument has a detection requirement below \qty{1}{\pico\mol\per\square\milli\metre} analyte (SNR $\geq$3) \citep{Goesmann2017}. However, this detection limit in \unit{\femto\mol\per\square\milli\metre} typically cannot directly be compared to mission requirements. To contextualise, the Europa Lander Science Definition Report specifies a LOD requirement of \qty{1}{\pico\mol} in a \qty{1}{\gram} (\qty{1}{\milli\litre}) water-ice sample \citep{Hand2017}. This is equivalent to \qty{141}{\femto\mol\per\square\milli\metre} if the \qty{1}{\gram} water-ice sample is deposited per cavity. If \qty{1}{\gram} of ice is deposited into a \qty{7.1}{\square\milli\metre} cavity via solvent evaporation, the $\text{LOD}_{3\sigma}$ for adenine of \qty{52}{\femto\mol\per\square\milli\metre} corresponds to approximately \qty{0.37}{\pico\mol} per gram of ice sample. This translates to a $\text{LOD}_{3\sigma}$ of \qty{0.050}{ppb} by mass for adenine in a \qty{1}{\gram} sample, compared to the requirement of \qty{0.135}{ppb}.

Detection limits were also determined for amino acids, lipids, and PAHs, all based on surface concentration. The detection limits for amino acids ranged between \qtyrange[range-phrase=~and~]{1}{1000}{\femto\mol\per\square\milli\metre} (\qtyrange{7.1}{7100}{\femto\mol\per\gram_{ice}}), with many below \qty{100}{\femto\mol\per\square\milli\metre}, though there was considerable variation between compounds. This variability aligns with expectations, as different compounds exhibit differing laser desorption and ionisation efficiencies, leading to variations in SNR. The LOD for three lipids, 17$\alpha$-ethynylestradiol, $\alpha$-tocopherol, and phylloquinone, was found to be below \qty{100}{\femto\mol\per\square\milli\metre} (\qty{710}{\femto\mol\per\gram_{ice}}). Similar observations were made for PAHs, with LODs in the range of tens of \unit{\femto\mol\per\square\milli\metre}. It is crucial to note that the LOD is compound-dependent and influenced by factors such as detector voltage, the number of sampled positions, and laser pulse energy. The detection sensitivity of ORIGIN can be further increased and different strategies, such as detector voltages, are discussed in more detail in Boeren et al. 2025.

These findings highlight ORIGIN’s potential as a powerful tool for detecting and quantifying a diverse range of organic molecules at trace abundances in planetary exploration contexts, offering new opportunities for identifying biosignatures in extraterrestrial environments.

\subsection{Network and Unsupervised Machine Learning Analysis}

Agnostic life detection helps reduce bias toward Earth-like biology. Therefore, sophisticated data analysis methods are required to confidently distinguish between abiotic and biotic signatures. Here, we present three possible pathways based on unsupervised machine learning (ML) and network analysis to increase the chances of successful life detection and improve space mission return. Applying unsupervised ML to mass spectrometric data helps analyse sample homogeneity using techniques like dimensionality reduction and clustering to identify patterns, similar mass spectra, and outliers containing trace element information. Networks, in turn, help study relationships within a dataset.

\begin{figure}[t]
\sidecaption[t]
\includegraphics[width = 6 cm]{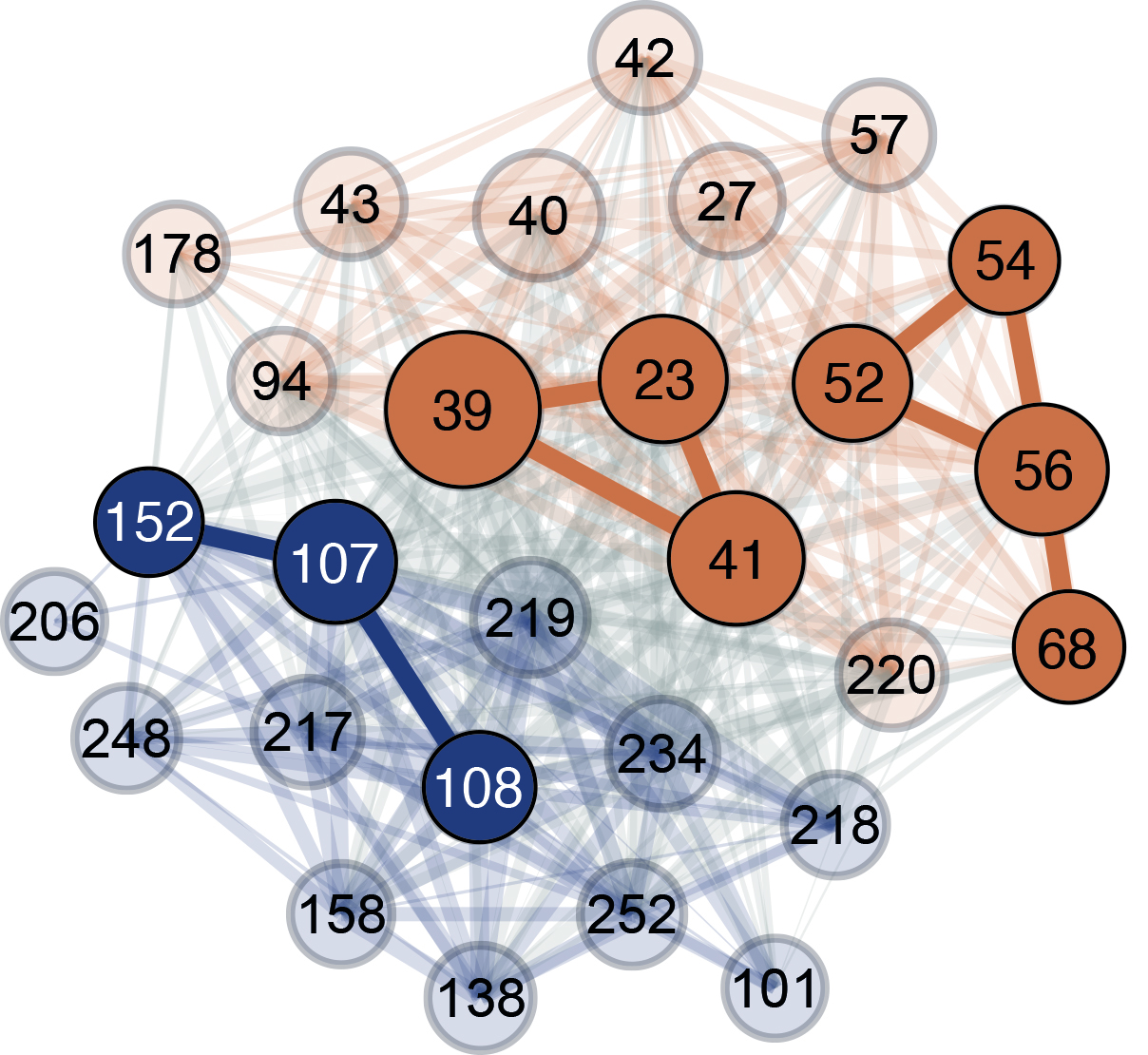}
\caption{Mass correlation network of a tyrosine – soil mixture using a threshold of \num{>0.85}. Blue nodes: amino acid masses. Orange nodes: soil and sample holder masses. Colour and saturation indicate cluster association, and node size reflects peak intensity in the spectra \citep{Schwander2022}. Figure adapted from \citep{Schwander2022}, licensed under CC-BY 4.0.}
\label{fig:9}     
\end{figure}

In general, networks consist of nodes that can be connected via edges. One example of a network applicable for mass spectrometric data is a so-called mass correlation network, a weighted, un-directed network where every node represents a certain mass value. For every pair of masses in the dataset, the Pearson correlation coefficients, $\rho$, are calculated, and the corresponding nodes are connected if the correlation exceeds a certain threshold. This allows to identify masses that belong to the same compound as they show a higher correlation, which is especially helpful when analysing mixture data where different compounds co-appear in a single mass spectrum. An example for this is shown in Fig. \ref{fig:9}, reproduced from \citep{Schwander2022}. ORIGIN was used to study a mixture of soil and the amino acid tyrosine, using a NIST reference soil, a Mars analogue, and a tyrosine standard. Three clusters, namely a tyrosine cluster, a soil cluster, and a sample holder and soil cluster were found. This method allows to assign mass spectra even from complex mass spectra to distinct chemical entities, and therefore presents an unbiased way to analyse mixture data and untangle peak patterns \citep{Schwander2022}. 

A different type of network is a nearest neighbor network or spectral similarity network. Whereas before in Fig. \ref{fig:9} a network node represented a mass value connected to other masses with high correlations, in a spectral similarity network each node represents a mass spectrum connected to other mass spectra with a certain degree of similarity, based on a selected metric. To effectively assess the similarity between mass spectra, dimensionality reduction techniques are especially useful to process the data prior to generating the network. These methods project high dimensional data, e.g., data containing a broad range of mass values (features), into a low-dimensional space while retaining as much of the original relationships between datapoints as possible. Datapoints, i.e., mass spectra, that have high similarity in their original feature space should remain similar in the low-dimensional space, which expresses itself in them being embedded close to each other.

\begin{figure}[!h]
\centering
\includegraphics[width = 11.7 cm]{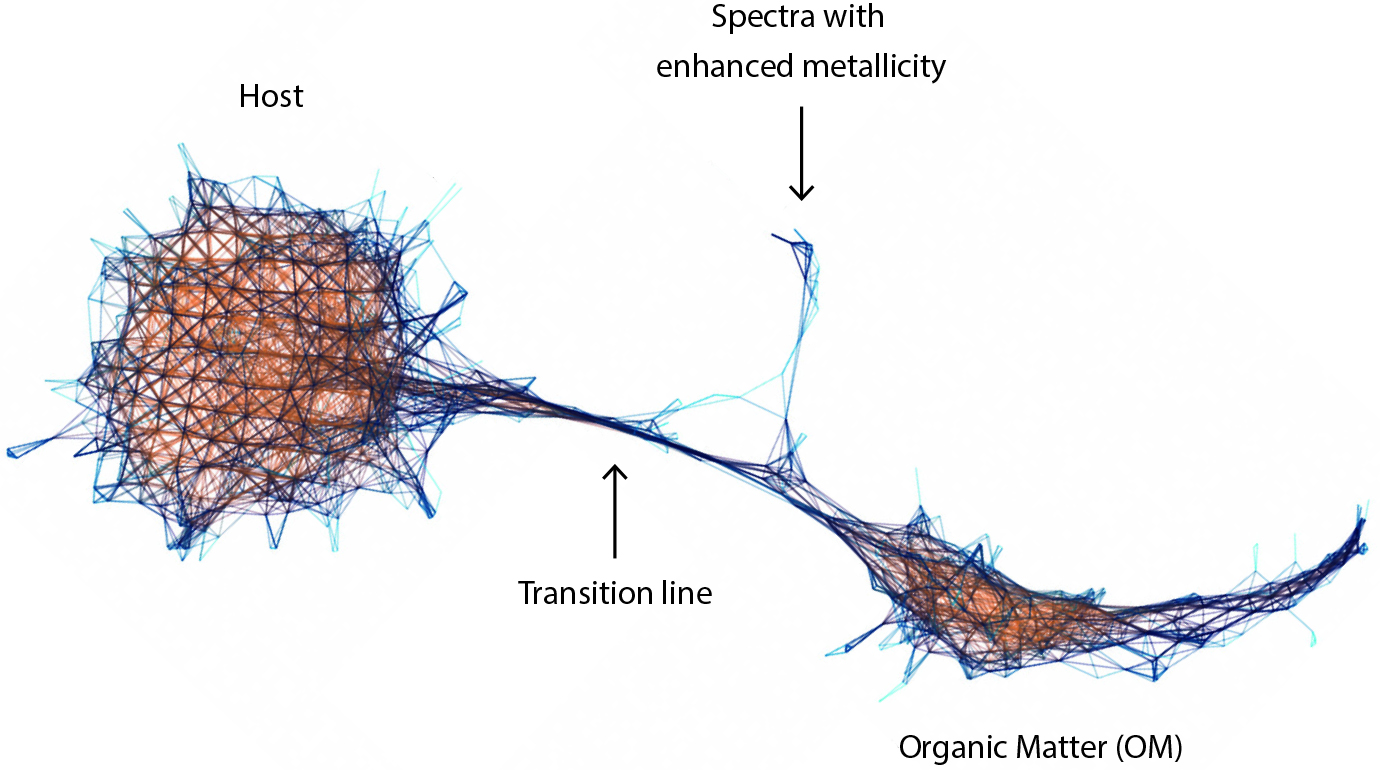}
\caption{Spectral similarity network of the Gunflint chert, reproduced from \citep{Lukmanov2022}, licensed under CC-BY 4.0. Two major clusters, corresponding to the host and the organic matter (OM) from the microfossils, can be identified. In addition, spectra with enhanced metallicity were found.}
\label{fig:10}     
\end{figure}

An analysis procedure based on dimensionality reduction and network analysis was applied to a 1.88 Ga Gunflint chert sample from Ontario, Canada, containing different species of microfossils \citep{Lukmanov2021}. Data were collected using the LMS instrument. Two dimensionality reduction techniques, namely principal component analysis (PCA) and uniform manifold approximation and projection (UMAP \citep{McInnes2018a, McInnes2018b}) were applied to the dataset to project it into a lower dimensional space. Using weighted mass correlation networks and PCA it was possible to identify chemical compositions in microfossils. UMAP enhanced this capability by differentiating chemical and mineralogical phases in the low-dimensional embeddings. When applied to data collected with a laboratory-scale LIMS instrument, UMAP successfully distinguished organic matter from inorganic host material \citep{Lukmanov2022}. 

Low-dimensional embeddings group points of high spectral similarity close to each other. Connecting every data point in the low-dimensional embedding with a certain number of nearest neighbours over an edge, a spectral similarity network can be drawn, as visualised in Fig. \ref{fig:10} for the Gunflint data. Two large clusters were found corresponding to the host material (chert) and the organic matter (OM) from the microfossils, respectively. The two clusters are connected through a transition line, corresponding to mass spectra that contained both signatures from the host and the OM. In addition, a few mass spectra with enhanced metallicity were found with the corresponding mass spectra forming outliers in the network, as their peak patterns differ both from the host and the OM.  

The example of the Gunflint dataset highlights the potential of dimensionality reduction techniques for data compression. Once a low-dimensional embedding has been obtained, a clustering algorithm can be used to identify groups of similar mass spectra. A general overview over the sample composition can then be gained by analysing just a few mass spectra from each cluster. In the example shown above in Fig. 10, comparing a mass spectrum from the host to a mass spectrum containing microfossil signatures would allow to see how they differ in composition and why they were embedded in different clusters. 

\begin{figure}[htbp]
\centering
\includegraphics[width = 11.7 cm]{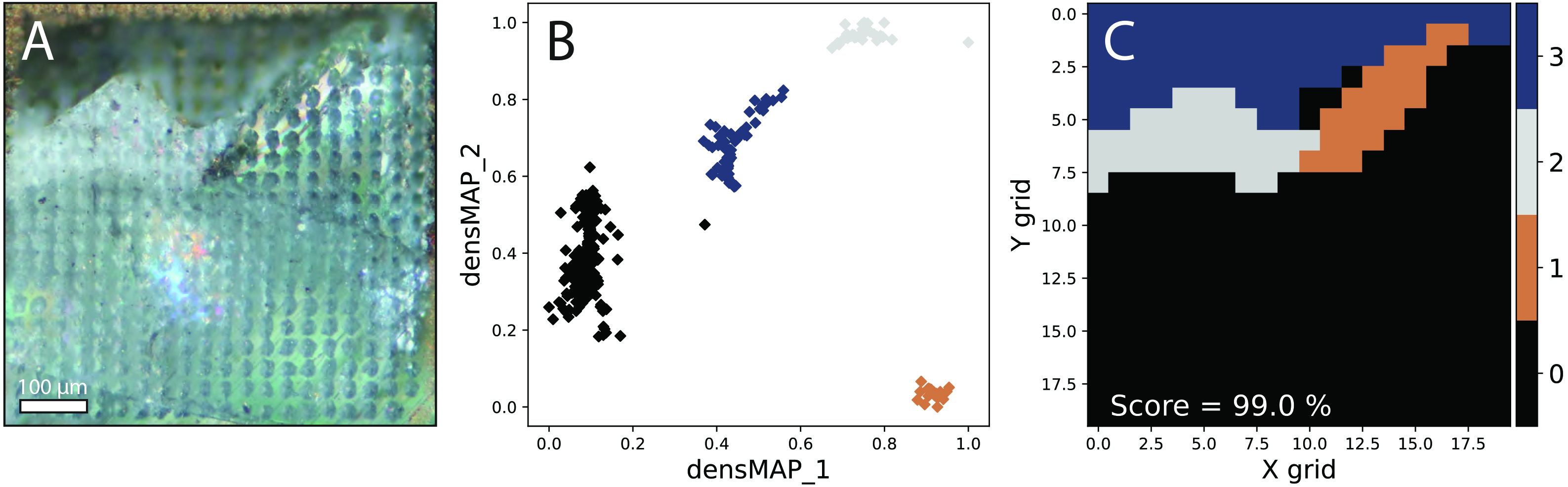}
\caption{Unsupervised analysis of a fluorapatite thin section. Panel A shows the microscopy image of the sampled area using LMS. Data collected from 400 locations were embedded into a two-dimensional space using densMAP, and subsequently clustered using hdbscan (panel B). In panel C the sampled area is coloured according to the retrieved cluster labels. Figure adapted from \citep{Gruchola2024}, licensed under CC-BY 4.0.}
\label{fig:11}     
\end{figure}

Recent developments, such as densMAP \citep{Narayan2021} — an extension of UMAP that preserves density — have enhanced the ability to detect minor mineral phases and trace elements, while also improving the handling of outliers in complex datasets. Fig. \ref{fig:11} visualises the analysis of a geological thin section using densMAP. Already from the microscopy image shown in panel A it is visible that different mineral phases are present. Using densMAP to embed the data recorded from 400 locations into a two-dimensional space results in four separate groups, as visualised in panel B. Using hierarchical density based clustering (hdbscan) \citep{McInnes2017} to find clusters within the embedding, the sampled area can be coloured according to the retrieved cluster labels for each location (panel C). The found pattern agrees well with the expected one from the microscopy image in panel A. A few datapoints in the embedding shown in panel B are only loosely connected to their cluster, with the corresponding mass spectra containing signatures of minor mineral phases, including pyrite, rutile, baddeleyite, and uranothorianite \citep{Gruchola2024}. To maximise the effectiveness of such methods, careful data pre-processing, including noise reduction and normalisation, is essential to minimise the impact of noise and improve overall data quality \citep{Gruchola2024}.

Such applications highlight the potential of unsupervised ML and network analysis to identify biosignatures and mineralogical features in extraterrestrial materials. Especially unsupervised ML techniques such as dimensionality reduction and clustering offer an autonomous pathway to group different spectral signatures present in a dataset, allowing for an unbiased detection of interesting features. This is especially crucial for agnostic life detection, as these methods distinguish between abiotic and biotic signatures independent of whether they are similar to signatures known from Earth or not.

\section{Ongoing mission preparation activities}
\subsection{Detection of Organics on Mars In-Situ}

Mars is one of the most promising objects in our Solar System where signatures of past or present life could be found in its protected subsurface environment. Consequently, as our close neighbour, Mars is one of the most visited targets in our Solar System in the search for signatures of life. For a few years now, we have been involved in the mission preparation for Abzu \citep{Wilhelm2021}, a proposed lander mission led by NASA Ames Research Center. Abzu will follow a novel approach to the search for signatures of life in the Martian subsurface: it focusses on the detection of lipids, which are robust organic molecules that could survive billions of years \citep{Aerts2014}. The lander will be equipped with the instruments Extractor for Chemical Analysis of Lipid Biomarkers in Regolith (ExCALiBR, NASA ARC)\citep{Wilhelm2024}, Merlin (a Raman spectrometer), and the Mars ORGanics ANAlyzer (MORGANA) \citep{Boeren2022}. ExCALiBR can digest up to 100 cm\textsuperscript{3} of soil material, grinding it, and subsequently provides a concentrated organic eluent through sophisticated solvent extraction protocols for chemical composition analysis. The processed sample is first analysed non-destructively by Raman spectroscopy and then destructively by mass spectrometry. The concept of ExCALiBR, by providing an eluent on a surface, closely aligns with the operation of ORIGIN, and due to the current measurement capabilities of ORIGIN, the system has been selected to obtain the seat of MORGANA.

\subsection{Icy Moons of Jupiter and Saturn}

The liquid oceans of Europa and Enceladus, moons of Jupiter and Saturn, respectively, represent additional promising subsurface habitats in our Solar System that could support life \citep{MacKenzie2021, Hand2017, Lunine2017}. Through the observed cracks in the icy sheets, water, potentially containing life signatures, can escape from the liquid oceans and form deposits on the surface. These surface deposits may be easier to target with landed missions than to sample the oceans by penetrator technologies. With the potential Orbilander to Enceladus \citep{MacKenzie2021} and Europa lander missions \citep{Hand2017}, there is strong interest from the astrobiology community in finding signatures of life, with the Orbilander mission favored by the recent NASA decadal survey \citep{NASA2023}. ORIGIN already meets many of the scientific requirements for organics detection for both the Enceladus Orbilander and Europa lander.

\subsection{Venus Atmosphere}

The middle and lower cloud deck of the Venusian atmosphere may provide another, more exotic habitat for life \citep{Seager2020, Morowitz1967, Cockell1999}. This cloud layer has a temperature of about \qty{60}{\degreeCelsius} and a pressure of about \qty{1}{\bar}, which are known to be more favourable conditions for life than the ones on the surface of Venus, with temperatures exceeding \qty{400}{\degreeCelsius} and tens of bars \citep{Seager2020}. The existence of a biosphere could explain some observations, such as UV absorption and abundances of molecular oxygen, which cannot be explained otherwise \citep{Zahn1985}. The Morning Star Mission (MSM) program (formerly known as the Venus Life Finder program), focuses on the detection and identification of signatures of life within the atmosphere of Venus, with a probe mission scheduled for launch in 2026, followed by a habitability and a sample return mission \citep{Seager2022a, Seager2022b}. Due to the measurement capabilities of ORIGIN \citep{Ligterink2022}, the MSM team expressed a strong interest in using this instrument for in-situ organic signatures detection and chemical composition analysis on the habitability mission, consisting of a balloon mission that will dive into the atmosphere \citep{Buchanan2022}.

\subsection{Future Data Analysis Pipelines}

The possibility of biosignatures being only sparsely present, if present at all, on other planetary bodies and moons must be considered. Having sophisticated data pre-processing techniques can help to increase the significance of a detection or a non-detection of signatures of life. This can be achieved using unsupervised ML methods based on dimensionality reduction, clustering, or autoencoders. Applying such methods directly on-board of a spacecraft bears the potential to separate data of interest containing signatures worth investigating from bulk data.

\begin{figure}[htbp]
\centering
\includegraphics[width = 11.7 cm]{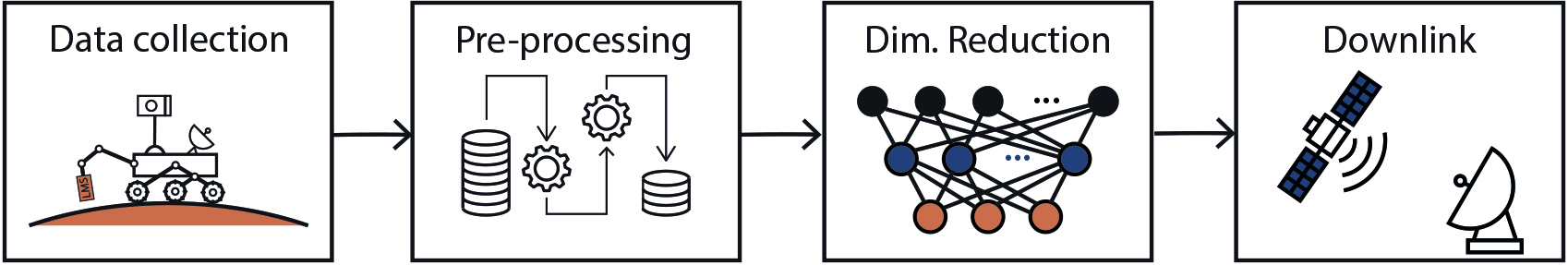}
\caption{Data compression using dimensionality reduction on board a spacecraft prior to data downlink. Figure adapted from \citep{Gruchola2024}, licensed under CC-BY 4.0.}
\label{fig:12}     
\end{figure}

A procedure such as that shown in Fig. \ref{fig:12} can be used to pre-process and reduce the dimensionality of data collected, for example, on a rover using a mass spectrometer or similar instrument, before using a selective downlink for representative data \citep{Gruchola2024}. Through this, data of interest, e.g., data containing signatures of potential biotic origin, could be determined, allowing to download the corresponding raw data selectively. This enables more efficient use of limited resources such as on-board storage and downlink rates. Data of little interest, e.g., from a mineral host or matrix with no interesting signatures, could be discarded on board the spacecraft, freeing up storage space and allowing more data to be collected. This increases the chance of a positive biosignature detection \citep{Gruchola2024}.

\section{Conclusions and Future Work}

The existence of extant or extinct life beyond Earth is one of humanity's most fundamental questions, driving space science and exploration for decades. Several space exploration missions are planned to visit different objects in our Solar System in the next two decades, from Venus to the outer Solar System, either to better understand the habitability conditions or to search for signatures of life. In this contribution, we highlighted the current measurement capabilities of our LIMS systems for the detection of different classes of signatures of life. We have demonstrated that this technology is capable of localising and analysing the chemical composition of microstructures in analogue material, including individual microbes and fossil structures with dimensions at the micrometre level. Additionally, it is capable of sulphur isotope fractionation measurements with an accuracy in the order of $\delta^{34}\text{S} = \qtyrange[range-phrase=-]{2}{3}{\permille}$, and the detection of various classes of organic molecules, with detection sensitivities at the level of \unit{\femto\mol\per\square\milli\metre}. The recently introduced data analysis approaches based on network analysis and unsupervised machine learning algorithms represent a major step forward in unbiased data analysis of mass spectrometric information for life signature detection. For example, clustering can be used to distinguish relevant from irrelevant data, which is beneficial for space exploration missions typically limited in available bandwidth, thus improving the overall scientific return of a mission.

The ORIGIN system is currently being redesigned to enable its deployment with ExCALiBR in the Mojave Desert in 2026 and 2027, with a focus on organics detection, in preparation for a future Mars exploration mission. In parallel, the measurement capabilities of ORIGIN for the detection of organics in the context of Venusian atmosphere will be further explored by investigating the implications for organics detection in a high sulphur environment. In addition, a novel measurement protocol will be developed to enable ORIGIN to measure organics from ice samples. Proof-of-concept measurements will be carried out on a glacier in the Swiss Alps.

\begin{acknowledgement}
The authors acknowledge the support from the Swiss National Science Foundation. This work has been carried out within the framework of the NCCR PlanetS supported by the Swiss National Science Foundation under grants 51NF40\textunderscore182901 and 51NF40\textunderscore205606.
\end{acknowledgement}

\ethics{Competing Interests}
{The authors have no relevant financial or non-financial interests to disclose.}

\bibliographystyle{styles/spphys_ed}
\bibliography{SSR2025.bib}
\end{document}